\documentclass[%
 reprint,
 amsmath,amssymb,
 aps,
aps,prx,
]{revtex4-2}

\usepackage{graphicx} 
\usepackage{dcolumn}  
\usepackage{bm}       
\usepackage{verbatim} 
\usepackage{gensymb}
\usepackage{lineno}   
\usepackage{soul}     %
\usepackage{color}
\usepackage{textcomp}
\usepackage{CJK}

\begin{document}

\preprint{APS/123-QED}

\title{Eckart streaming with nonlinear high-order harmonics: an example at gigahertz }

\author{Shiyu Li}
\affiliation{State Key Laboratory of Ocean Engineering, School of Ocean and Civil Engineering, Shanghai Jiao Tong University, Shanghai, 200240, China}%
\author{Weiwei Cui}
\affiliation{State Key Laboratory of Precision Measuring Technology and Instruments, Tianjin University, Tianjin 300072, China}
\author{Thierry  Baasch}
\affiliation{Lund University}
\author{Bin Wang}
\affiliation{State Key Laboratory of Ocean Engineering, School of Ocean and Civil Engineering, Shanghai Jiao Tong University, Shanghai, 200240, China}
\affiliation{Key Laboratory of Marine Intelligent Equipment and System, Ministry of Education, China}
\author{Zhixiong Gong}%
\email{Corresponding author: zhixiong.gong@sjtu.edu.cn; URL: https://en.naoce.sjtu.edu.cn/teachers/GongZhixiong.html}
\affiliation{State Key Laboratory of Ocean Engineering, School of Ocean and Civil Engineering, Shanghai Jiao Tong University, Shanghai, 200240, China}
\affiliation{Key Laboratory of Marine Intelligent Equipment and System, Ministry of Education, China}%

\date{\today}

\begin{abstract}
Acoustic streaming shows great potential in applications such as bubble dynamics, cell aggregation, and nano-sized particle isolation in the biomedical and drug industries.
As the acoustic shock distance decreases with the increase of incident frequency, the nonlinear propagation effect will play a role in acoustic streaming, e.g., Eckart (bulk) streaming at a few gigahertz (GHz).
However, the theory of source terms of bulk streaming is still missing at this stage when high-order acoustic harmonics play a role.
In this paper, we derive the source term including the contribution of higher-order harmonics.
The streaming-induced hydrodynamic flow is assumed to be incompressible and no shock wave occurs during the nonlinear acoustic propagation as restricted by the traditional Goldberg number $\Gamma < 1$ or $\Gamma \approx 1$ which indicates the importance of nonlinearity relative to dissipation.
The derived force terms allow evaluating bulk streaming with high-order harmonics at GHz and provide an exact expression compared to the existing empirical formulas. 
Numerical results show that the contribution of higher-order harmonics increases the streaming flow velocity by more than  $20 \%$. 
We show that the expression introduced by Nyborg should be avoided in numerical computations as it includes part of the acoustic radiation force that does not lead to acoustic streaming. 
\end{abstract}

\pacs{Valid PACS appear here}
\maketitle

\section{\label{sec:Introduction}Introduction}
Gigahertz (GHz) acoustics have recently been used in experiments for nanoparticle trapping, enrichment, and separation based on the acoustic streaming effect \cite{cui2016localized,cui2019trapping,yang2022self}.
Acoustic manipulation in the gigahertz range has potential applications including nano-sized biosensors \cite{cui2019trapping}, nanoliter microreactors \cite{shilton2014nanoliter}, and microfluid jet producer \cite{dentry2014frequency}.
However, the study of GHz streaming is just at the beginning due to the challenges of fabrication techniques of ultrahigh frequency resonators \cite{cui2019theoretical} and the huge computational costs with direct numerical simulations at such small wavelengths especially in three dimensions \cite{Steckel2021thesis}. 
For a typical GHz tweezer, the wavelength is 1.5 $\mu$m at the frequency of 1 GHz in water, which is much smaller than typical microchannel sizes, e.g., a few tens or hundreds of micrometers.
In addition, it is easy to induce nonlinear propagation at GHz since the acoustic shock distance depends on the working frequency and the large vibration velocity on the transducer surface.
These make the theoretical and experimental studies of GHz acoustical tweezers more challenging the same technique in the frequency regime of megahertz (MHz).

The sound waves are produced based on mechanical vibration and are coupled into the fluid medium for microfluidic applications. The highest efficiency of energy conversion from electrical to mechanical energies typically occurs at the mechanical resonance, which depends on the desired acoustic wavelength (or driving frequency).
The wavelengths of typical GHz transducers in the coupling medium range from around 100 nanometers to a few micrometers, leading to the difficulties of device fabrication. 
Typical piezoelectric transducers (PZT) use the vibration of planar sources to produce acoustics with a frequency regime of 1 - 10 MHz in the field of acoustofluidics. 
However, it is a challenge to fabricate PZT at the thickness of nanometers for the frequency at GHz since the resonance depends on the selected vibration mode depending on the piezo thickness.
Considering interdigitated transducers \cite{white1965direct} at GHz, the distances between electrode fingers are too small to fabricate with the commonly-used fabrication process and cannot withstand high power \cite{cui2019theoretical}.
This is partly solved with the successful fabrication of the high-tone bulk acoustic resonators on four substrates with very high Q factor (up to 48000) at 1GHz \cite{zhang2006high}.
Then, Cui \textit{et al.} combined the fabrication of FBAR (Film Bulk Acoustic Wave Resonator) technique with micro/nanofluidics and developed the field of GHz acoustofluidics \cite{cui2016localized}.

Compared with the recent development of fabrication techniques and experimental demonstrations of GHz acoustical tweezers, the theories of acoustic bulk streaming at such high frequencies are not well studied. 
For most of the published experimental works at GHz, empirical expressions of the source term of acoustic streaming are used in the numerical simulations, and no one seems to verify the streaming flow velocities between the simulation and experiment results \cite{cui2019trapping,yang2022self}. 
Indeed, it is easy to understand that the streaming-induced flow patterns are similar in the confined microchannels because of the mass conservation of the steady flow.
Since the seminal works of Eckart \cite{eckart1948vortices}, Nyborg \cite{nyborg1953acoustic,Nyborg1965BookChapter} and Lighthill \cite{lighthill1978acoustic}, the systematic theories of the bulk (Eckart) streaming are built and developed. A good historical perspective of bulk acoustic streaming can be found in Ref. [\onlinecite{baudoin2020acoustic}].
It should be noteworthy that Nyborg derived an expression of the source term of bulk streaming to solve the Stokes equation for the hydrodynamic flow velocity \cite{nyborg1953acoustic,Nyborg1965BookChapter} which is widely used in the research community of this field.  However, as pointed out by Lighthill, the source term by Nyborg contains a gradient term that makes a contribution to the acoustic radiation pressure instead of streaming \cite{lighthill1978acoustic,baudoin2020acoustic}. The source term of acoustic streaming was recast recently by Riaud \textit{et al.} specifying the sole source of bulk streaming without the gradient of acoustic Lagrangian \cite{riaud2017influence}. 
In their work, they consider the bulk streaming inside sessile droplets of size 1 mm under the activation of surface acoustic waves at a frequency of around 20 MHz neglecting the nonlinear propagation since Gol’dberg number [defined in Eq. (\ref{Eq.A1: Goldberg number}) below] is much smaller than 1 and the droplet size is much smaller than the shock wave distance. Hence, the source term is limited to linear propagation when there are no high-order acoustic harmonics. 
However, the nonlinear effect of acoustic propagation can not be neglected in the frequency regime of GHz since the viscous dissipation is remarkable and the dimensionless Gol’dberg number is comparable to the unit. This is the case for the recent experimental works at GHz \cite{cui2016localized}.

In this work, we revisit the source term of bulk acoustic streaming with two assumptions: (i) acoustics are rotational, and (ii) the streaming-induced steady hydrodynamic flow is incompressible. 
Only weakly nonlinear effects with harmonic waves are taken into consideration and the shock wave is outside of scope since it will make the second assumption fail.
Both theoretical and numerical examples are proposed to illustrate the issue of Nyborg's expression which should be avoided as shown in section \ref{sec3:Fluid mechanics and source term}. More importantly, based on the peculiar characteristics of bulk streaming at GHz, the nonlinear effect of acoustic propagation will be considered and a theoretical source term for this situation is provided in terms of pressure fields of different orders of harmonics. This work provides a theoretical basis for steady streaming at GHz with high-order harmonics.
\begin{figure}[!htbp]
\includegraphics[width=8.0cm]{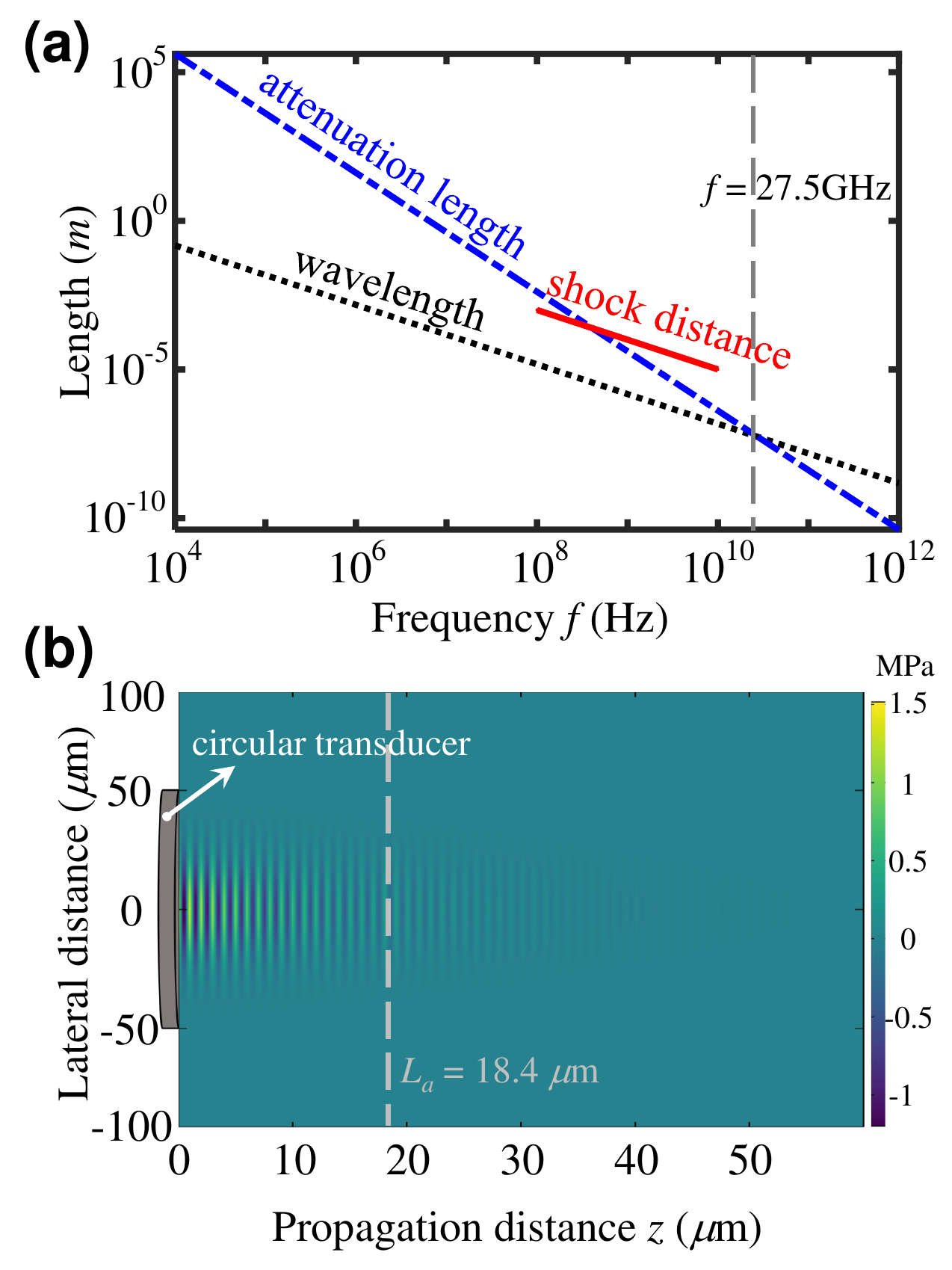}
\caption{\label{Fig1: Length vs frequency} (Color online) (a) The wavelength, acoustic attenuation length, and shock distance versus the incident frequency from 0.01 MHz to 1000 GHz (both axes are in logarithmic coordinates). $B/A=$ 5.1 is taken from Ref. [\onlinecite{sapozhnikov2021review}] for water. The vibration velocity to estimate the shock distance is $U_{ac}$ = 1 m/s based on previous experimental results from 0.1 to 10 GHz. The attenuation length $L_{a}$ is equal to the incident wavelength $\lambda=$ 0.055 $\mu$m at $f \approx 27.5$ GHz (see the grey dashed line). For a special case at frequency $f = 1.5$ GHz, the wavelength is $\lambda=$ 1 $\mu$m, the attenuation length $L_a = 18.4$ $\mu$m, and the shock distance is $L_s = $ 67.2 $\mu$m. The pressure field in the propagation plane with a circular transducer at $f=$ 1.5 GHz is shown in (b).}
\end{figure}

\section{\label{sec2:Different lengths up to GHz} Characteristic lengths in the frequency regime near 1 GHz}
\begin{table}[!htbp]
\small
  \caption{ Physical parameters. The fluid medium in this work is water. $f$ is the incident frequency. $A=\rho_0 c_0^2$ and B are two classical acoustic coefficients in nonlinear acoustics. Another nonlinear parameter is defined as $\beta = 1+B/2A$. $U_{ac}$ is the magnitude of the acoustic velocity perturbation at the transducer surface. Note that a factor of 2 is missing in the definition of $L_a$ in Refs. [\onlinecite{baudoin2020acoustic}] and [\onlinecite{riaud2017influence}].} 
  \label{Table 1 Acoustic properties}
  \begin{tabular*}{0.48\textwidth}{@{\extracolsep{\fill}}lll}
    \hline 
    Symbol & Physical parameter      & Value \\
    \hline 
    $\rho_0$   & static density      & 1000 kg/m$^3$    \\
    $c_0$      & acoustic velocity   & 1500 m/s   \\
    $\omega$   & angular frequency   & $2 \pi f$    \\
    $\mu_s$    & dynamic viscosity   & $1.002 \times 10^{-3}$ Pa s  \\
    $\nu$      & kinematic viscosity & $\mu_s/\rho_0$    \\
    $\mu_b$    & bulk viscosity      & $2.8 \times 10^{-3} $ Pa s \\
    $b$        & defined coefficients& $4/3 + \mu_b/\mu_s$ \\
    \hline
    $\Gamma$   & Goldberg number    & $L_a/L_s$             \\
    $L_a$        & acoustic attenuation length     & $2 \rho_0 c_{0}^{3} / [\omega^2 \mu_s b ]$   \\
    $L_s$      &  acoustic shock distance        & $c_{0}^{2} / [\omega \beta U_{ac}]$     \\
    Re$_{ac}$  & acoustic Reynolds number   & $ L_a / \lambda $  \\
    Re$_{hd}$  & hydrodynamic Reynolds number   & $\rho_0 |\mathbf{v_2}| L_c/\mu_s$  \\
    $\delta$      &  boundary layer thickness            &  $ \sqrt{2 \nu / \omega}$     \\
    $L_c$      &  microchannel height            & $60 \, \mu$m     \\
    \hline
  \end{tabular*}
\end{table}

Compared with the general frequencies in the medical ultrasonic regime (e.g., 1-10 MHz), there are two different characteristics in the regime near 1 GHz as shown in Fig. \ref{Fig1: Length vs frequency}(a): (i) the attenuation length $L_a = 2 \rho_0 c_{0}^{3} / [\omega^2 \mu_s b]$ is just 1 order of magnitude as the wavelength at typical MHz regime, and the acoustic Reynolds number is $Re_{ac} = L_a/\lambda \gg 1$ \cite{riaud2017influence}. 
$\rho_0$ is the static mass density of the propagation medium, $c_0$ is the sound speed,  $\omega$ is the angular frequency, and $b=4/3 + \mu_b/\mu_s$ is the defined coefficient related to dynamic ($\mu_s$) and bulk viscosities ($\mu_b$) for convenience.
The related physical parameters and values for the medium of water are listed in Table \ref{Table 1 Acoustic properties}. 
At $f=1.5$ GHz in water, the wavelength is $\lambda=1$ $\mu$m and the attenuation length is $L_a=18.4$ $\mu$m [see Fig. \ref{Fig1: Length vs frequency}(b) for the pressure distribution]. While at $f=1.5$ MHz in water, it has $\lambda=1$ mm and $L_a=1.84 \times 10^4$ mm. 
(ii) the attenuation length is comparable with the shock distance $L_s = c_{0}^{2} / [\omega \beta U_{ac}]$ at the typical vibration velocity of the transducer working at GHz (e.g., $U_{ac} = 1$ m/s). $\beta = 1+ B/[2A]$ is the nonlinear parameter with two nonlinear acoustic coefficients $A$ and $B$.
At $f= 27.5$ GHz as marked with the grey dashed line in Fig. \ref{Fig1: Length vs frequency}(a), the attenuation length equals the wavelength, i.e., $L_a = \lambda = 0.055$ $\mu$m. That is to say, most of the energy from the transducer will dissipate in the propagation distance of one wavelength. 
In this work, only low GHz transducers for acoustic steaming will be studied and they have been used most in recent microfluidics experiments in the regime of GHz. 
In fact, the shock distance describes the nonlinearity of the acoustic propagation and the attenuation length is an indicator of the wave dissipation in the medium.
Here, the Goldberg number $\Gamma$ is introduced to provide a dimensionless measure of the importance of nonlinearity relative to dissipation with the definition
\begin{equation}
\begin{aligned}
\Gamma = \frac{L_a}{L_s} = \frac{2 \rho_0 c_0 \beta U_{ac}}{\omega \mu_s b}
\label{Eq.A1: Goldberg number}
\end{aligned}
\end{equation}
Note that the attenuation length $L_a$ is independent of the activation velocity $U_{ac}$, while the shock distance $L_s$ has a linear relation with it.
In general, there are no high-order harmonics during the acoustic propagation with $\Gamma \ll 1$ which is the case in Ref. [\onlinecite{riaud2017influence}]. When $\Gamma$ is around the unit, it is possible to induce the high-order harmonics which will be discussed in detail in Sec. \ref{sec4: acoustic harmonics}. 
Note that the shock distance depends on the $U_{ac}$ and for the case of $U_{ac} = 1$ m/s in Fig. \ref{Fig1: Length vs frequency}(a), $\Gamma $ equals 1 (i.e., $L_s = L_a$) at $f= 0.407 $ GHz. However, the source term of bulk streaming with the consideration of high-order acoustic harmonics is not available and this will be solved in the present work.

\section{\label{sec3:Fluid mechanics and source term} Revisit of source term for bulk streaming}
\subsection{\label{sec3A:source term of bulk streaming}Source term of bulk streaming}
Before studying the streaming effect with high-order harmonics, we need to revisit the source term of bulk streaming. By following the work in Ref. [\onlinecite{riaud2017influence}], we briefly recall the source term for the Eckart-type acoustic streaming with bulk waves. To derive the final formula of the streaming source term, we start with the constitutive equations including the conservation of mass and momentum in the fluid medium as

\begin{equation}
\begin{aligned}
&\frac{\partial \rho}{\partial t}+\nabla \cdot(\rho \mathbf{v})=0 \label{Eq. 1: mass conservation}
\end{aligned}
\end{equation}
\begin{equation}
\begin{aligned}
&\frac{\partial \rho \mathbf{v}}{\partial t}+\nabla \cdot(\rho \mathbf{v} \otimes \mathbf{v}) = -\nabla p+\mu_{s} \Delta \mathbf{v}+\left(\frac{\mu_{s}}{3}+\mu_{b}\right) \nabla \nabla \cdot \mathbf{v} \label{Eq.2: momentum conservation}
\end{aligned}
\end{equation}
with $\rho$, $\mathbf{v}$, and $p$ are the density, velocity vector, and pressure, respectively. $t$ designates the time. The entropy ($s$) balance is ensured for the system with $\mathrm{d} s=0$ and the state equation is 
\begin{equation}
\begin{aligned}
&p=p(\rho), \text { with }\left.\frac{\partial p}{\partial \rho}\right|_{s}=c_{0}^{2} \label{Eq.3: state equation}
\end{aligned}
\end{equation}
The first-order equation of state is easily obtained by using the Taylor expansion with $p_1 = c_0^2 \rho_1$, where $p_1$, $\rho_1$ are the first-order acoustic pressure and density.
In general, compared with the viscous effect, the thermal effect is negligible because it is proportional to $\gamma - 1$ with $\gamma$ the adiabatic index which is weak in liquids \cite{riaud2017influence}. Since only small perturbation occurs with respect to the hydrostatic parameters, we can apply the perturbation method and hence decompose the physical field $\mathcal{X}$ into three parts: hydrostatic $\mathcal{X}_{0}$, acoustic $\mathcal{X}_{1}$, and hydrodynamic $\mathcal{X}_{2}$.
Under the assumption with the perturbation method, we assume that $\mathcal{X}_{0}\gg{\mathcal{X}_1}\gg{\mathcal{X}_2}.$
$\mathcal{X}$ can be either a scalar or a vector.
The time average of acoustic component $\mathcal{X}_{1}$ is equal to zero, written as $\left<\mathcal{X}_{1}\right> = 0$, while the hydrodynamic part $\mathcal{X}_{2}$ shows the nonlinear feature with its time average not equal to zero (i.e., $\left<\mathcal{X}_{2}\right> \neq 0$). Based on the above assumptions, we can expand the density $\rho$, pressure $p$, and velocity $\mathbf{v}$ as follows
\begin{equation}
\begin{aligned}
&\rho=\rho_{0}+{   \rho}_1+{  \rho}_2, \\
&p=p_{0}+   {p}_1+  {p}_2 \\
&\mathbf{v}=\mathbf{0}+   {\mathbf{v}}_1+  {\mathbf{v}}_2 .
\label{Eq.4: second-order expansion}
\end{aligned}
\end{equation}
Here the particle velocity $\mathbf{v_0}$ is zero when not disturbed by acoustic waves. 
By expanding the above mass and momentum conservation equations up to the first- and second-order with some algebraic operations [see details in Appendix \ref{Appendix A}], the source term of acoustic streaming can be derived as 
\begin{equation}
\mathbf{F_s} = - \left(\frac{4 \mu_s} {3}+\mu_b \right) \left\langle \frac{\rho_1}{\rho_0} \Delta \mathbf{v}_1\right\rangle,
\label{Eq.5: source term with harmonics}
\end{equation}
where $\Delta$ is the Laplace operator with $\Delta \mathbf{v}_1 = \nabla ^2 \mathbf{v}_1 = \nabla \cdot \nabla \mathbf{v}_1$. 
It is noteworthy that this formula excludes the contribution of the gradient of acoustic Lagrangian, which in fact will not induce the bulk acoustic streaming \cite{lighthill1978acoustic,riaud2017influence}. In addition, the gradient of acoustic Lagrangian is very large compared to the sole source term which is possible to result in large computational errors. The effect of the acoustic Lagrangian gradient has not been studied before and will be demonstrated numerically in this work [see Sec. \ref{sec3C: 2D case}].

As shown in Appendix \ref{Appendix B1} by using the linear wave equation for mono-chromatic waves, the source term for acoustic streaming in Eq. \eqref{Eq.5: source term with harmonics} can be simplified as 
\begin{equation}
\mathbf{F_s} = \left(\frac{4 \mu_s}{3}+\mu_b \right) \frac{\omega^{2}}{\rho_{0} c_0^{4}}\langle {p_1 \mathbf{v}_1}\rangle 
\label{Eq. 7: single frequency source term}
\end{equation}
which depends on the time-averaged acoustic intensity $\langle {\mathbf{I}}\rangle =\langle {p_1 \mathbf{v}_1}\rangle$. It indicates that once the acoustic field is calculated, the source term to compute the steady fluid streaming ($\mathbf{v_2}$) is available to solve the Stokes equation derived in Eq. \eqref{Eq. A12: Stokes equation of the streaming flow}
\begin{equation}
-\nabla p^*_2+\mu_s \Delta \mathbf{v}_2 + \mathbf{F_s} = 0
    \label{Eq. Add1: Stokes equation of the streaming flow}
\end{equation}
where $p^*_2$ is the modified hydrodynamic pressure to remove the contribution of the average acoustic Lagrangian.

It is noteworthy that the classical source term for acoustic streaming simulation proposed by Nyborg \cite{nyborg1953acoustic, Nyborg1965BookChapter} is widely used although it should be avoided: 
\begin{equation}
\mathbf{F_s}^{Nb}=-\rho_{0} \nabla \cdot\langle{\mathbf{v_1}} \otimes {\mathbf{v_1}}\rangle
\label{Eq.9:Nyborg expression}
\end{equation}
where $\langle\cdot\rangle$ is the time average operator. Recent studies have shown that this expression is not completely related to acoustic streaming, which may cause large numerical errors, and the source term can be divided into two parts through mathematical derivation \cite{riaud2017influence,baudoin2020acoustic}.
In the following, we propose an analytical example of a one-dimensional (1D) bulk standing wave and a numerical simulation of a two-dimensional (2D) traveling wave to show the difference between the streaming field induced by these two source terms, i.e., based on Eq. \eqref{Eq. 7: single frequency source term} and Eq. \eqref{Eq.9:Nyborg expression} by Nyborg.

\begin{figure}[!htbp]
\includegraphics[width=6.6cm]{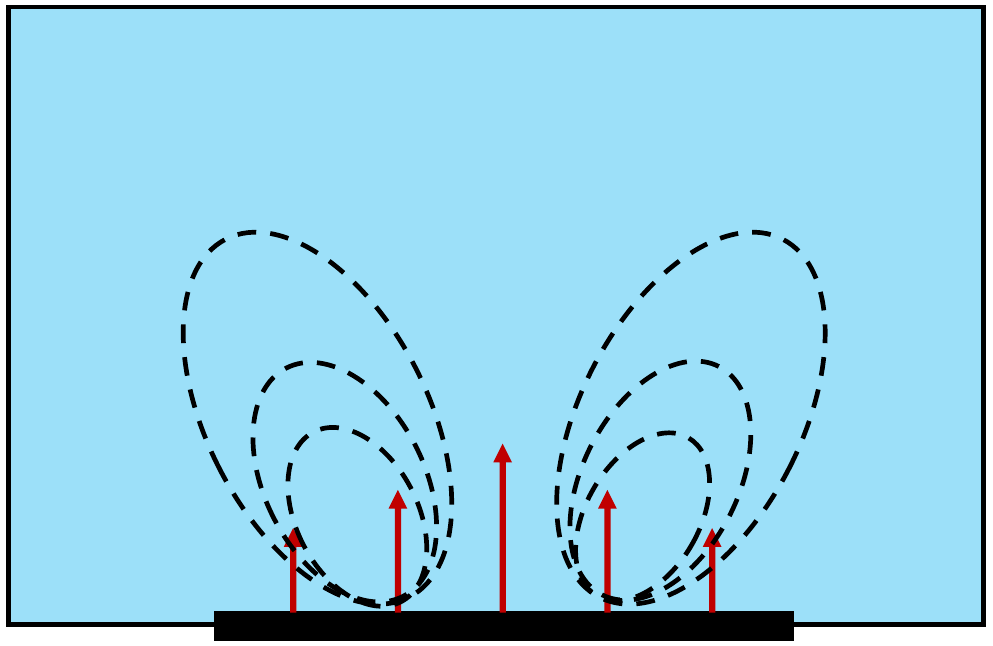}
\caption{\label{Fig2:Skematic of streaming} (color online) The schematic of a 2D acoustic streaming by a planar transducer. The resonator (black area) excites GHz plane waves in the fluid (blue) and forms vortices (closed dashed lines) under the activation of body force (red arrow). In the acoustic simulation, the wall is set as the impedance boundary condition, and in the flow field simulation, the wall is set as the no-slip condition. 
Based on the conservation of fluid inside the cavity, the Eckart streaming always pushes the fluid up in the center and rolls back again from the two sides.}
\end{figure}

\subsection{\label{sec3B: 1D bulk standing wave example} One-dimensional bulk standing wave example}
In this section, we take the ideal 1D bulk standing waves as an example to show the difference of the source terms from Nyborg and in Eq. \eqref{Eq. 7: single frequency source term}. This simple case will easily show the contradiction of the two source terms for bulk streaming and can be well understood from the point of view of streaming physics.
The acoustic velocity field and pressure field of the 1D plane standing waves can be expressed as the addition of two ideal counter-propagating plane waves:
\begin{subequations}
\begin{equation}
\mathbf{v_{1}} = 2 v_{am} \sin (k z) \cos (\omega t) \mathbf{e_z}
\end{equation}
\begin{equation}
p_{1} = 2 p_{am} \cos (k z) \sin (\omega t)
\end{equation}
\end{subequations}
where $v_{am}$ and $p_{am}$ are the amplitudes of the acoustic velocity and pressure, $k$ is the wavenumber, $z$ is the space coordinate with the unit vector in the propagation direction $\mathbf{e_z}$.
By substituting them into the Nyborg source term in Eq. \eqref{Eq.9:Nyborg expression}, we can get the body force in the propagation direction as:
\begin{equation}
\mathbf{F_s}^{Nb} = -2 \rho_{0} k v_{am}^{2} \sin (2 k z) \left\langle \cos (2\omega t) +1 \right\rangle
\label{standingwaveforce}
\end{equation}
Obviously, the magnitude of the force $\mathbf{F_s}^{Nb}$ does not vanish since the time dependence $ \left\langle \cos (2\omega t) +1 \right\rangle $ is not equal to 0. On the other hand, we can get the following different result if we use the source term in Eq. (\ref{Eq. 7: single frequency source term}):
\begin{equation}
 \mathbf{F_s}  = \left(\frac{4 \mu_s}{3}+\mu_b \right) \frac{\omega^2}{c_0^4 \rho_0}\left\langle p_1 \mathbf{v}_1\right\rangle \propto \left\langle \sin(\omega t) \cos(\omega t)\right\rangle
 \label{body force SW this paper}
\end{equation}
which vanishes as 0 after the time average procedure in one period. It is clear that these two results are contradictory.
\begin{figure}[!htbp]
\includegraphics[width=8.6cm]{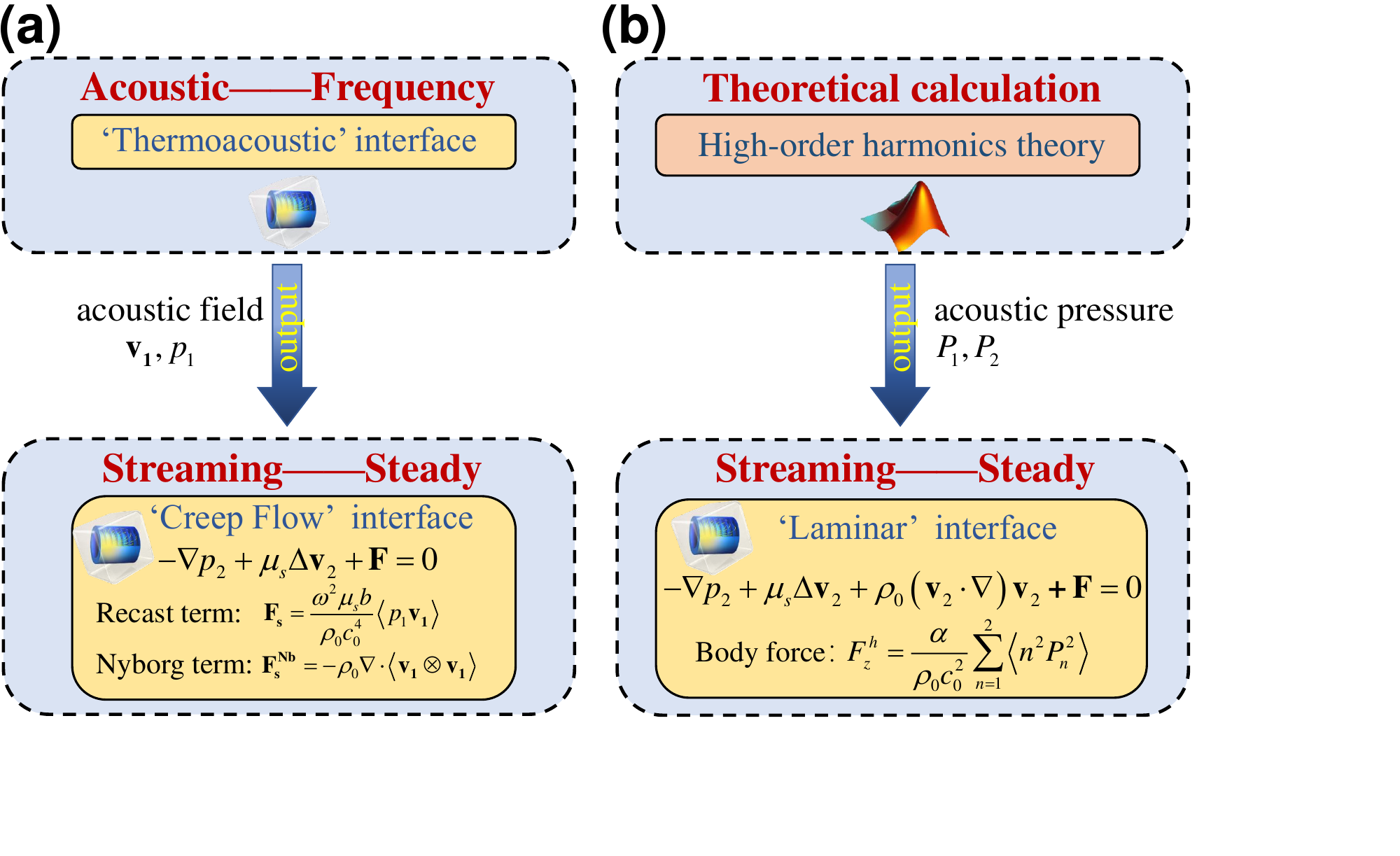}
\caption{\label{Fig3: computational flowchart} (Color online)  The flowcharts of the simulation for bulk streaming (a) when the hydrodynamic Reynolds number is much smaller than 1; (b) when nonlinear propagation is considered.}
\label{structure}
\end{figure}
As observed from Eq. (\ref{Eq. 7: single frequency source term}), the body force depends on the average acoustic intensity which is uniform in space for progressive ideal plane waves. Since the standing waves can be regarded as the addition of two counter-propagating plane waves, the body forces of each plane wave should have the same magnitude while in reversed directions, leading to the null of the total body force for the streaming effect. This physical explanation agrees with the results of Eq. (\ref{body force SW this paper}).
Indeed, the source term $\mathbf{F_s}^{Nb}$ by Nyborg is not entirely related to the streaming, and it contains the gradient of acoustic Lagrangian which contributes to acoustic radiation pressure instead of streaming \cite{lighthill1978acoustic,riaud2017influence}.

\subsection{\label{sec3C: 2D case} Numerical examples of 2D bulk streaming }
Since the wavelength at GHz is in the order of nanometers or a few microns, it often needs small size steps in the computational domain based on the rules of thumb. Hence, the three-dimensional simulation of bulk streaming problems will suffer the challenge of a massive amount of numerical computation cost. To reduce the numerical burden, we build a 2D model as illustrated in Fig. \ref{Fig2:Skematic of streaming}, and this will not hinder our understanding of the acoustic streaming mechanism at this stage. 
\begin{figure} [!htbp]
\includegraphics[width=8.6cm]{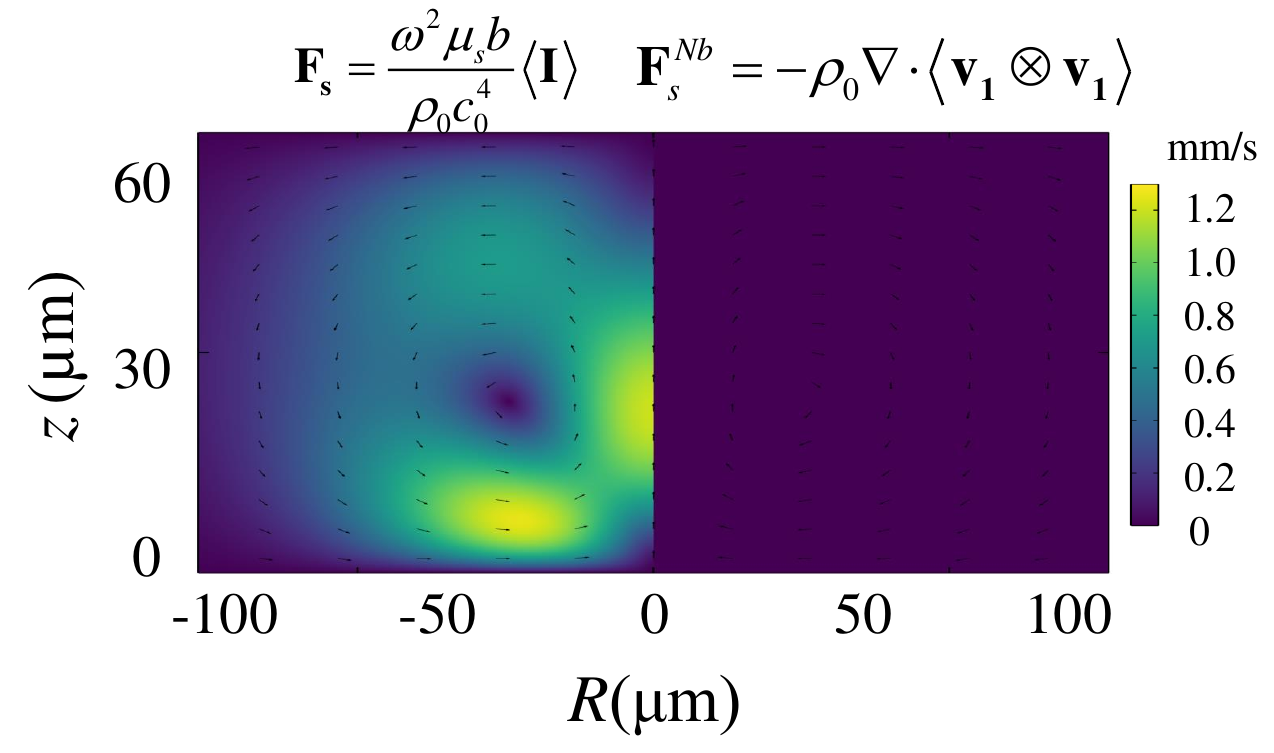} 
\caption{(Color online) The streaming simulation of two source terms at 1.5 GHz when the vibration velocity of the transducer surface is $U_{ac} = 0.1 $m/s. The radius of the transducer is $a=50$ $\mu$m. The channel height is $L_c = 60$ $\mu$m and the width is 200 $\mu$m. The left half is simulated by using the recast source term [Eq.\eqref{Eq. 7: single frequency source term}], and the right half is simulated with Nyborg's source term [Eq.\eqref{Eq.9:Nyborg expression}]. The uniform colorbar is used. The color background represents the velocity distribution of the hydrodynamic flow and the arrows represent the flow directions. Significant differences between the two source terms are observed to calculate the streaming-induced velocities. The maximum flow velocity by Nyborg's expression is 0.02 mm/s, which is much smaller than that with the present source term in Eq. (\ref{Eq. 7: single frequency source term}).}
\label{Fig4: Sketch}
\end{figure}
The contradiction between the two streaming source terms has been revealed by the ideal 1D standing wave model in Sec.\ref{sec3B: 1D bulk standing wave example}. This section will further show the difference of hydrodynamic flow velocities through numerical simulations. 
In the following simulations, only the source terms are different [i.e., Eqs. (\ref{Eq. 7: single frequency source term}) and (\ref{Eq.9:Nyborg expression})].
We use water as the propagation medium with the parameters listed in Table \ref{Table 1 Acoustic properties}. The excitation frequency is $f=1.5$ GHz and the vibration velocity of the transducer with a radius 50 $\mu$m is $U_{ac} = $ 0.1 m/s, leading to the Goldberg number $\Gamma = 0.027 \ll 1$. Under this condition, the source term of streaming in Eq. (\ref{Eq. 7: single frequency source term}) is suitable without the consideration of nonlinear propagation.
The boundary layer thickness (or viscous penetration depth) is $\delta = \sqrt{2 \nu / \omega}=$ 22 nm $\ll L_c = 60$ $\mu$m with $L_c$ the height of the microchannel.
Meanwhile, the microchannel size is smaller than the shock distance $L_s = 67.2$ $\mu$m so that the shock waves will not be accumulated and formed. Hence, the bulk streaming is dominant in the fluid domain and the Rayleigh boundary streaming can be negligible \cite{vanneste2011streaming}. 

The numerical simulations of acoustic streaming in this paper are carried out by COMSOL Multiphysics 6.0 \cite{COMSOL6.0} with the flowchart shown in Fig. \ref{structure}(a).
For the calculation of the first-order acoustic field, we use the ``Thermoacoustic'' Interface. The oscillation velocity boundary is used to replace the effect of the resonator, and the impedance boundary condition is adopted for the walls. 
Then, the steady flow simulations are conducted based on the ``creep flow" interface with the input of the source terms under the circumstance that the hydrodynamic Reynolds number is $\text{Re}_{hd}= 0.072 \ll 1$. All the walls were set with no-slip boundary conditions. 
Indeed, the simulation with the ``laminar" interface obtains the same streaming results as the ``creep flow" interface case (not shown in the following for brevity).

\begin{figure} [!htbp]
\includegraphics[width=8.6cm]{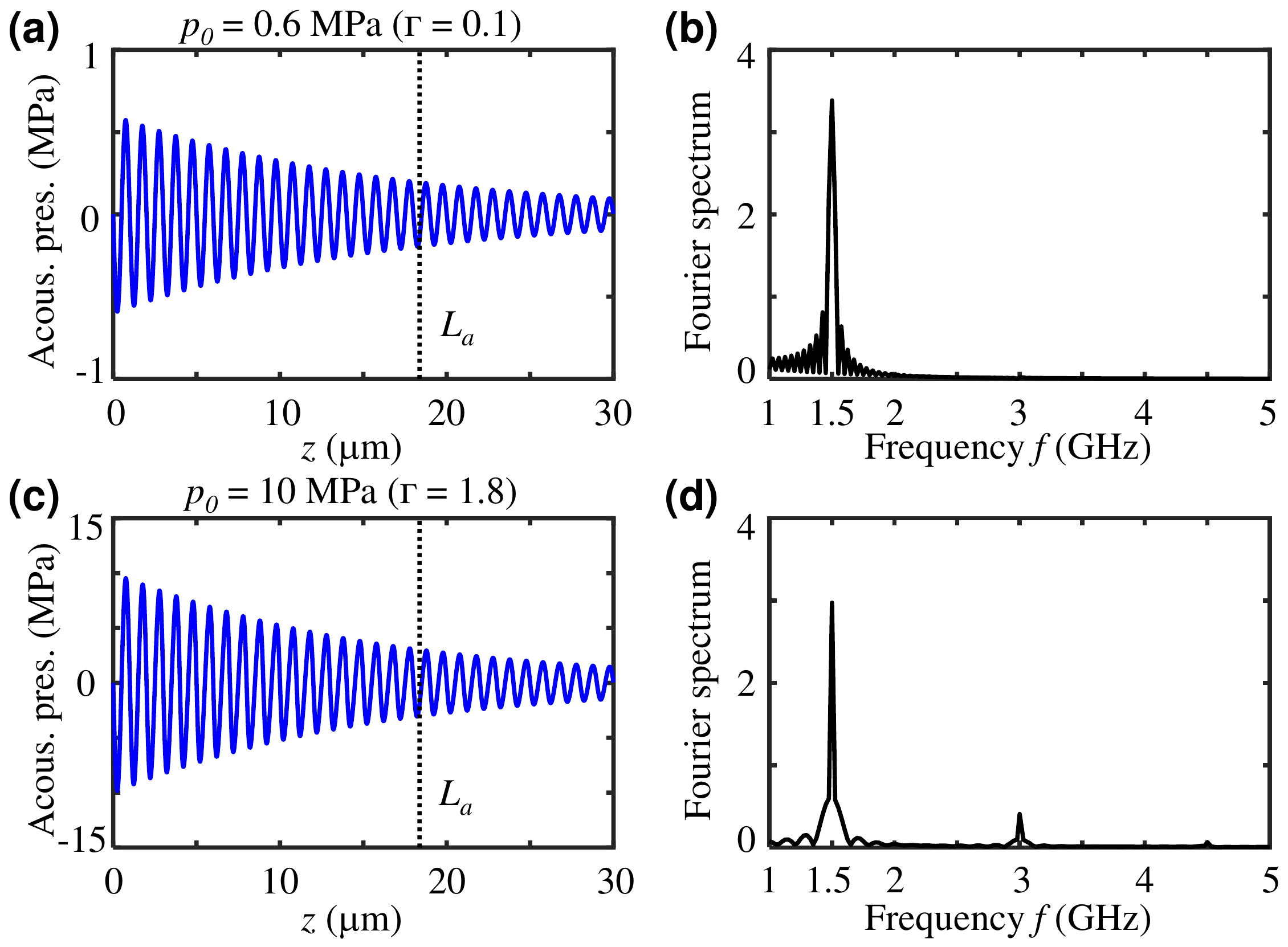} 
\caption{\label{Fig5: pressure and frequency spectrum} (color online) Acoustic pressure based on the theoretical method in Ref. [\onlinecite{Du1986Harmonic}] and frequency spectrum with two different Goldberg numbers. 
    (a) For $\Gamma = 0.1$, the pressure amplitude is $p_{am} = 0.6$ MPa with the attenuation distance $L_a=18.4$ um (indicated by the black dotted line) and the shock distance $L_s=168.1$ um. 
    (b) Only the fundamental wave appears along with a few sidelobes. There are no obvious high-order harmonics in the frequency spectrum.
    (c) For $\Gamma = 1.8$, the pressure amplitude $p_{am} = 10$ MPa with $L_a=18.4$ um and $L_s=10.1$ um. 
    (d) Different from the results in (b), the second-order harmonic wave appears in this case.}
\end{figure}

In the 2D numerical simulations of bulk streaming, we set the volume forces in the propagation direction $F_z$ and the lateral direction $F_r$ with the flow velocities indicated by the colormap of the background and the directions indicated by the black arrows in Fig. \ref{Fig4: Sketch}. 
Because of the symmetry of the hydrodynamic flows in the microchannel, we show half of the simulation results and put them together for ease of comparison: the left half is for the source term in Eq. (\ref{Eq. 7: single frequency source term}) while the right half is for the source term by Nyborg in Eq. (\ref{Eq.9:Nyborg expression}).
It is obvious that the streaming velocities of these two terms have a significant difference at this time. Note that the main differences lie in the magnitudes of the maximum flow velocities, while the hydrodynamic flow patterns are similar because of the conservation of fluid flow in the microchannels.
Through this example, it is obvious that there are problems in using the classical source term by Nyborg for acoustic streaming simulation. 
In addition, it will be noteworthy that an empirical expression of the source term \cite{cui2016localized,yang2022self} is widely used for the present simulations of GHz streaming which is limited to the plane wave case and could not predict the exact values of the hydrodynamic flow velocities.

\section{\label{sec4: acoustic harmonics} Weakly nonlinear propagation with high-order acoustic harmonics}
The source term of the bulk streaming in Eq. (\ref{Eq. 7: single frequency source term}) is proper under the assumption of linear acoustic propagation.  
However, it does not apply if nonlinear acoustics are considered (e.g., high-order harmonics) for the streaming phenomenon since the pressure field may include the contribution of multiple frequencies. 
To solve the issue, we re-derive a general expression of the sole source term for acoustic streaming with the multi-frequency acoustic field in terms of only acoustic pressure $p_1$ as follows [see Eq. \eqref{Eq. B3: source in terms of only pressure}]
\begin{equation}
\mathbf{F_s} = \left(\frac{4}{3} \mu_s+\mu_b\right) \left\langle\frac{{p}_{1}}{c_{0}^{4} \rho_{0}^{2}} \nabla \frac{\partial {p}_{1}}{\partial t}\right\rangle 
\label{Eq. 13: source in terms of only pressure}
\end{equation}
With the detailed derivation of the source expression given in Appendix \ref{Appendix B2}. This formula has the advantage to compute the incompressible hydrodynamic fluid motion induced by acoustic streaming once the acoustic field can be calculated. 

Recall the Goldberg number as first introduced in Sec. \ref{sec2:Different lengths up to GHz}, which plays an important role in the viscous process considering nonlinear effects \cite{Blackstock1964Thermoviscous}. It measures the relative importance of nonlinear effects and dissipation effects. Since only weakly nonlinear propagation is considered in this work and there are no shock waves as assumed, the following will derive the source terms with  Goldberg numbers within $\Gamma \ll 1$ and $\Gamma \sim 1$ based on the analytical expressions of harmonic waves from finite-amplitude Gaussian beam in a fluid by Du and Breazeale \cite{Du1986Harmonic}. 

\subsection{Acoustic pressure and frequency spectrum at different $\Gamma$}\label{sec4B: no harmonics}
To calculate the acoustic streaming with nonlinear propagation, we need to solve the pressure field according to Eq.(\ref{Eq. 13: source in terms of only pressure}).
The nonlinear propagation of acoustics could be calculated with either analytical methods or numerical simulations.  
This part will review the analytical expressions of the acoustic pressure fields of the fundamental and second-order harmonics by Du \& Breazeale for a Gaussian beam from a piston transducer \cite{Du1986Harmonic}.
As Goldberg points out, the shock wave will not be formed when $\Gamma < 1$. This is because the dissipation effect is significant under the circumstance, and the harmonics cannot be effectively accumulated. For the piston transducer, we assume that the vibration velocity of the sound source satisfies the Gaussian function and take the Gaussian coefficient as a unit for simplicity:
\begin{equation}
U(R)=U_{ac} e^{-R^{2} / a^{2}}
\end{equation}
where $R$ is the radial coordinate, and $a$ is the radius of the piston transducer. By taking the method from Du and Breazeale, we can expand its acoustic pressure waveform into Fourier series (see Eqs. (A1) and (A4) in Ref. [\onlinecite{Du1986Harmonic}]):
\begin{equation}
p_{1}=\sum_{n=1}^{2} P_{n} \sin \left[n \omega\left(t-\frac{z}{c_{0}}\right)\right]
\label{harmonic}
\end{equation}
where $P_{n}$ represents the spatial components of the $n_{th}$ harmonic acoustic pressure, $t$ designates time, and $z$ is the distance from the transducer surface in the propagation direction. This expression illustrates the physical mechanism of the nonlinear pressure field in terms of the addition of different orders of harmonics. For the fundamental component (the first order)
\begin{equation}
P_{1}(f \mid z)=p_{am} e^{-\alpha z} \exp \left(-R^{2} / a^{2}\right) \label{simplify P1}
\end{equation}
and for the second-order harmonics
\begin{equation}
\begin{split}
P_{2}(f \mid z)=\frac{p_{am}^{2} k \beta e^{-4 \alpha z}}{4 \alpha \rho_{0} c_{0}^{2}} e^{\left(-2 R^{2} / a^{2}\right)} \left(e^{2 \alpha z}-1\right) 
\end{split}
\label{simplify P2}
\end{equation}
where the pressure amplitude at the center of the transducer surface is $p_{am}=\rho_{0}c_{0}U_{ac}$, and $\alpha = 1/L_a$ is the attenuation coefficient.
The detailed derivation is briefly organized in Appendix \ref{Appendix B3}.

An example shows that the nonlinear effect starts around $\Gamma = 3$ for a plane wave by Hamilton and Blackstock, while other scholars select 4.5 as the threshold for shock formation \cite{hamilton2016effective}. 
To compare the pressure fields with or without high-order harmonics, two pressure amplitudes are selected to make the Gol’dber numbers $\Gamma = 0.1$ and $\Gamma = 1.8$ for the Gaussian beam, respectively. The radius of the piston transducer is $a=50$ $\mu$m with the excitation frequency $f= 1.5$ GHz.
The pressure fields along the propagation direction $z$ are computed with the analytical method and numerical simulations based on COMSOL \cite{nonlinearacoustic}. These two results agree well with each other and are shown (one blue curve for brevity) in Fig. \ref{Fig5: pressure and frequency spectrum}(a) at $p_{am} = 0.6$ MPa and (c) at $p_{am} = 10$ MPa. 
The attenuation distance $L_a$ is indicated by the vertical dashed lines. 
To illustrate the nonlinear high-order harmonics when $\Gamma > 1$, the frequency spectrum based on fast Fourier transform is calculated as shown in Fig. \ref{Fig5: pressure and frequency spectrum}(b) and (d).
It could be observed that there is a sharp peak at $f=3$ GHz for $\Gamma = 1.8$ in the frequency spectrum of Fig. \ref{Fig5: pressure and frequency spectrum}(d) which comes from the contribution of the second-order harmonics. 

\subsection{Source terms of streaming with and without high-order harmonics}\label{sec4C: with harmonics}
\begin{figure}[!htbp]
\includegraphics[width=7.6cm]{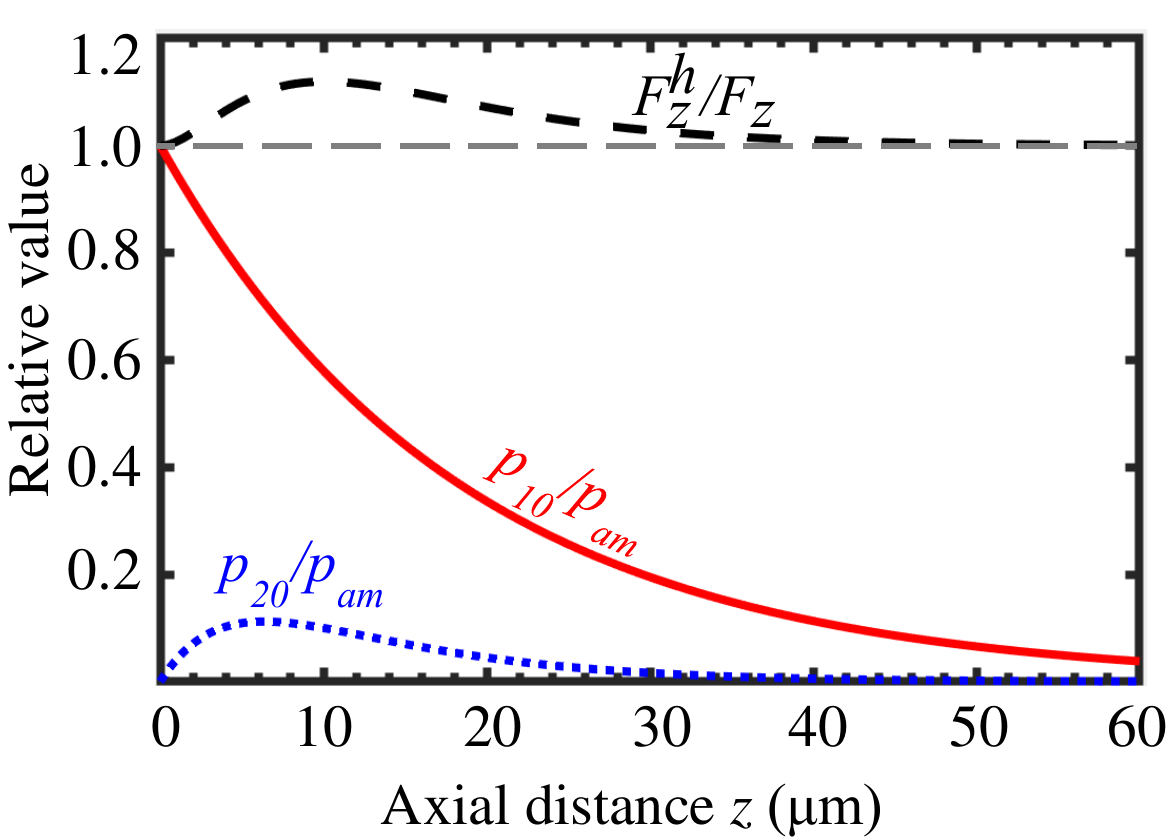}
\caption{\label{Fig6:harmonics curve} (Color online) The ratios of pressure magnitude of the fundamental wave $p_{10} / p_{am}$ (red solid line), second harmonic wave $p_{20} / p_{am}$ (blue dotted line), and the body force $F_z^h / F_z$ (black dashed line) change along the axial direction when the pressure magnitude is $p_{am} =$ 10 MPa and the frequency is 1.5 GHz. $F_{z}^{h}$ is the axial body force considering the contribution of both the fundamental and the second-order harmonics, which is defined by Eq. \eqref{harmonic force}. $F_z$ is the volume force considering only the fundamental frequency acoustic pressure, which is determined by Eq. \eqref{force without harmonic}. The fundamental sound pressure decreases with the increase of axial distance, and the second harmonic sound pressure and body force increase first and then reduce with the increase of axial distance. The attenuated energy of the fundamental wave turns into high-order harmonics.}
\end{figure}

According to the general source term of acoustic streaming in Eq. (\ref{Eq. 13: source in terms of only pressure}), explicit expressions with and without high-order harmonics could be obtained once the pressure fields are known. 
For the situation with only fundamental component, by the insertion of Eqs.  \eqref{harmonic} with \eqref{simplify P1} into Eq. \eqref{Eq. 13: source in terms of only pressure}, we can rewrite the axial body force $F_z^s$ as (note $n$ is truncated up to 1 for the fundamental wave case):
\begin{equation}
F_z^s = \frac{\alpha}{\rho_{0} c_{0}^{2}}\left| P_{1}\right|^{2} \label{force without harmonic}
\end{equation}
where
$\left|P_{1}\right|^{2}= p_{am}^{2} e^{-2 \alpha z} e^{-2 R^{2} / a^{2}}$. Similarly, by substituting Eq. \eqref{harmonic} into Eq. \eqref{Eq. 13: source in terms of only pressure}, the axial body force $F_{z}^{h}$ considering the second-order harmonics can be expressed as:
\begin{equation}
F_{z}^{h}=\frac{\alpha}{\rho_{0} c_{0}^{2}} \sum_{n=1}^{2}\left( n^{2} \left|P_{n}\right|^{2}\right) 
\label{harmonic force}
\end{equation}
where the upper right ``$h$'' represents the body force with the contribution of the second-order harmonics. Indeed, this source term works for high-orde harmonics when $n > 2$. When only the fundamental wave is considered ($n=1$), Eq. \eqref{harmonic force} degenerates into \eqref{force without harmonic}. Note that the lateral component of the source term vanishes after the time average procedures, \textit{i.e.}, $F_R = 0$ for the cases with and without high-order harmonics. 

To study the contribution of the second-order harmonics on the total pressure field and body force, the same piston transducer is used with the initial pressure amplitude $p_{am} = 10$ MPa at the excitation frequency $f = 1.5$ GHz. 
The sizes of the microchannel are the same as those in Fig. (\ref{Fig4: Sketch}).
The simulation flowchart is given in Fig. \ref{Fig3: computational flowchart}(b) with the pressure field computed based on the theoretical method in Matlab and the induced streaming obtained by using COMSOL. Note that the hydrodynamic flow speed can reach up to 1 m/s, leading to $Re_{hd} \gg 1$, the `laminar' interface is applied in this section.
The normalized pressure fields versus the propagation distance $z$ of the fundamental (red solid line) and second-order (blue dotted line) harmonic waves are shown in Fig. \ref{Fig6:harmonics curve}. The surface of the transducer is defined as $z =0$. The amplitude of the fundamental component decrease versus $z$ because of the wave absorption.
While the pressure amplitude of the second-order harmonics increases from the transducer surface to a maximum value at $z \approx 6$ $\mu$m, which agrees with the fact that the nonlinearity effect is dominant over the dissipation close to the source. 
In addition, the relative axial body force with the consideration of the second-order harmonics with respect to only the fundamental wave contribution is plotted versus $z$ with the black dashed line in Fig. \ref{Fig6:harmonics curve}. A horizontal grey dashed line is also provided as a reference. It is shown that the excess of the body force comes from the contribution of the second-order harmonics.

\begin{figure}[!htbp]
   \centerline{\includegraphics[width=8.6cm]{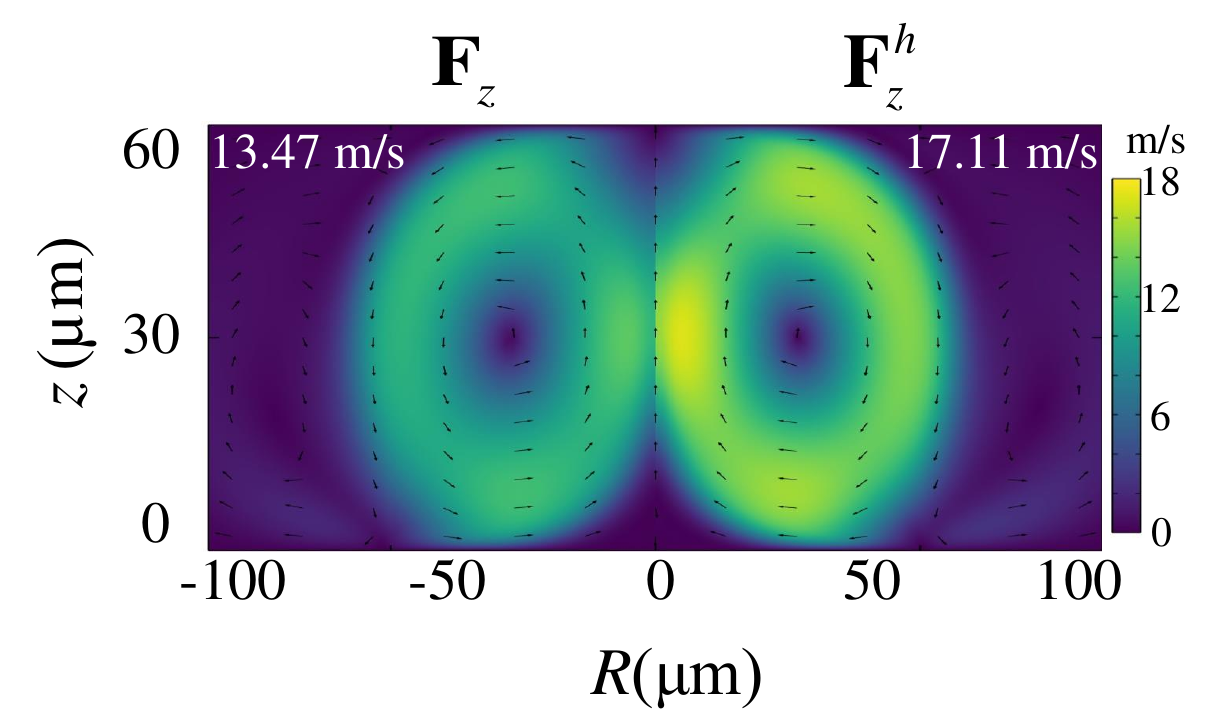}}
    \caption{\label{Fig7:without harmonics} (color online) Streaming results without and with the second-order harmonics at 1.5 GHz. The acoustic pressure is 20 MPa. The left half shows the acoustic streaming field with only the contribution of the fundamental wave with the maximum flow speed of 13.47 m/s, it is calculated by Eq.\eqref{force without harmonic}. The right half considers the contribution of the second-order harmonics with the maximum flow speed of 17.11 m/s, which is based on Eq.\eqref{harmonic force}. The color depth represents the streaming velocity.}
\end{figure}

To further explore the streaming-induced hydrodynamic motion in the microchannel, we take the same configuration as in Fig. \ref{Fig6:harmonics curve} but double the pressure amplitude $p_{ac} = 20$ MPa to enhance the contribution of the second-order harmonics.
The left half part of Fig. \ref{Fig7:without harmonics} shows the streaming patterns with only the fundamental component, while the right half is for the situation with both the fundamental and second-order harmonics.
The colormap in the background indicates the velocity amplitude of the hydrodynamic flows with the arrows giving the flow directions.
The maximum hydrodynamic flow velocity increases more than 20$\%$ if the second harmonic component is taken into consideration, as shown in Fig. \ref{Fig7:without harmonics}.

\section{\label{sec5: conclusion and discussion}Conclusion and discussion}
The theory of the source term for bulk acoustic streaming has been developed with an emphasis on the nonlinear propagation in the frequency regime of GHz.
A dimensionless number called Goldberg number ($\Gamma$) is introduced to measure the importance of acoustic nonlinearity relative to dissipation at GHz.
This work makes it possible to compute the streaming-induced microfluidic motion for the recent GHz devices in a better manner instead of using the empirical formulas \cite{cui2016localized,yang2022self}. As shown in the numerical simulation, the contribution of high-order harmonics could increase the maximum velocity of hydrodynamic flows up to 20\% if only the fundamental component is considered in linear propagation.
In addition, the source term of bulk streaming in linear acoustic propagation is revisited in the form of the wave damping and acoustic intensity \cite{riaud2017influence}.
More importantly, we propose analytical and numerical examples to show the contradictory results by the source term of linear propagation and the classic expression by Nyborg \cite{nyborg1953acoustic, Nyborg1965BookChapter}. The term by Nyborg should be avoided by others (e.g., see Ref. [\onlinecite{eisener2015characterization}]) since it contains the contribution of acoustic radiation pressure \cite{lighthill1978acoustic,riaud2017influence}.

The alternative method to compute the bulk streaming is the full-model direct numerical simulation (DNS) \cite{Steckel2021thesis}. However, it will take more computational cost and will be challenging to handle the full simulation in the 3D domain.
The computational burden could be relieved with the effective boundary conditions developed by Bach and Bruus \cite{bach2018theory} and using the axial symmetry of the geometrical model \cite{Jorgensen2022thesis}.
However, attention should be still paid to the simulation of 3D bulk streaming if there is no axial symmetry with either the present theory or DNS, for instance, the drop-shaped transducer working at GHz in a microchannel \cite{yang2022self}.
It should be noted that this work could be helpful to design acoustical tweezers for particle trapping with the consideration of both the acoustic radiation pressure\cite{silva2011expression,gong2021equivalence} and the streaming-induced drag force.

\begin{acknowledgments}
Z.Gong Thanks for the support from the National Natural Science Foundation of China (24Z990200542) and Shanghai Jiao Tong University for the startup funding.
\end{acknowledgments}

\appendix

\section{\label{Appendix A} Derivation of the sole source of acoustic streaming}
\subsection{\label{Appendix A1} First-order perturbation expansion}
To derive the source term of acoustic streaming (i.e., sound-induced fluid motion), we need to bridge the acoustic field (acoustic pressure $p_1$ and hydrodynamic velocity $\mathbf{v}_1$) with the fluid velocity $\mathbf{v}_2$. Using the perturbation method, we need to expand the fields up to the first-order and second-order terms. Firstly, we expand the fields up to the linear limit with the following perturbation
\begin{equation}
\begin{aligned}
&\rho=\rho_{0}+   \rho_{1}\\
&p=p_{0}+   p_{1} \\
&\mathbf{v}=\mathbf{0}+   \mathbf{v}_{1}
\label{Eq.A1: linear perturbation}
\end{aligned}
\end{equation}

Substituting Eq. \eqref{Eq.A1: linear perturbation} into \eqref{Eq. 1: mass conservation} and Eq. \eqref{Eq.2: momentum conservation}, we can obtain the first-order formulas of mass and momentum conservation respectively
\begin{equation}
\frac{\partial \rho_{1}}{\partial t}+\rho_{0} \nabla \cdot \mathbf{v}_{\mathbf{1}}=0 
\label{Eq. A2: first order mass conservation}
\end{equation}
\begin{equation}
\rho_{0} \frac{\partial \mathbf{v}_{1}}{\partial t} 
= -\nabla p_{1}+\mu_s \Delta \mathbf{v}_{1}+\left(\frac{\mu_s}{3}+\mu_b\right) \nabla \nabla \cdot \mathbf{v}_{1} 
\label{Eq. A3: first order momentum}
\end{equation}
where $\Delta \mathbf{v}_{1} = \nabla \cdot (\nabla\mathbf{v}_{1})$. Since the first-order acoustic field is defined as irrotational (i.e., $\nabla \times \mathbf{v}_1 =0$), Eq. \eqref{Eq. A3: first order momentum} can be further changed into the following form:

\begin{equation}
 \rho_{0} \frac{\partial \mathbf{v}_{1}}{\partial t}=-\nabla p_{1} +\left (\frac{4}{3}\mu_s+\mu_b\right) \Delta \mathbf{v_{1}} 
 \label{Eq. A4: first-order momentum conservation}
\end{equation}

It should be noted that the derivation of the above equation uses the mathematical identity of the vector Laplacian: $\nabla\nabla\cdot\mathbf{v_{1}}=\nabla^{2}\mathbf{v_{1}}+\nabla\times\nabla\times\mathbf{v_{1}}$ [see Eq.(3.70) in Ref. \onlinecite{arfken2013mathematical}].

\subsection{\label{Appendix A2} Second-order perturbation expansion}
Since acoustic streaming is a nonlinear phenomenon, we need to expand the fields up to the second-order as given in Eq. \eqref{Eq.4: second-order expansion}. After substitution into the mass conservation equation, one obtains
\begin{equation}
  \frac{\partial \rho_{2}}{\partial t} + \nabla \cdot (\rho_{1}\mathbf{v_{1}}) +\rho_{0}  \nabla \cdot \mathbf{v_{2}} = 0
\end{equation}

Before expand the momentum conservation equation [Eq. \eqref{Eq.2: momentum conservation}] up to the second-order, we first use the mass conservation to cancel some items in the momentum conservation [see Eq. \eqref{Eq.2: momentum conservation}]
\begin{equation}
\begin{split}
& \left[\frac{\partial\rho}{\partial t}+\nabla \cdot \left(\rho\mathbf{v}\right) \right] \mathbf{v} + \rho\frac{\partial \mathbf{v}}{\partial t}+\rho\mathbf{v}\cdot\nabla{\mathbf{v}} \\
& =-\nabla p+\mu_s \Delta\mathbf{v} + \left(\frac{\mu_s}{3}+\mu_b\right)\nabla\nabla\cdot\mathbf{v}
\label{Eq. A6: Momentum conservation}
\end{split}
\end{equation}
with the terms in the first square brackets vanishing based on the global mass conservation as given in Eq. \eqref{Eq. 1: mass conservation}.
Note that the identity $\nabla \cdot(\rho \mathbf{v} \otimes \mathbf{v}) = \rho \mathbf{v} \cdot \nabla(\mathbf{v})+\mathbf{v} \nabla \cdot(\rho \mathbf{v})$ is used here. By insertion of Eq. \eqref{Eq.4: second-order expansion} into \eqref{Eq. A6: Momentum conservation} and taking the equation up to the second order:
\begin{equation}
\begin{split}
& \rho_{0} \frac{\partial \mathbf{v_{2}}}{\partial t}+\rho_{1} \frac{\partial \mathbf{v_{1}}}{\partial t}+\rho_{0} \mathbf{v_{1}} \cdot \nabla \mathbf{v_{1}} \\
& =-\nabla p_{2}+\mu_s \Delta \mathbf{v_{2}} + \left (\frac{\mu_s}{3}+\mu_b \right) \nabla \nabla \cdot \mathbf{v_{2}}
\label{Eq. A7: second order Momentum Conservation}
\end{split}
\end{equation}

Since only the steady acoustic streaming is of interest, the first term of Eq. \eqref{Eq. A7: second order Momentum Conservation} vanishes with $\partial \mathbf{v_{2}}/ \partial t = 0$. Throughout the whole work, the incompressible flow assumption is hold, so that
\begin{equation}
\nabla \cdot \mathbf{v}_2=0
\label{Eq. A8: imcopressible fluid assumption}
\end{equation}
The last term on the right-hand side in the second-order momentum conservation equation [Eq.\eqref{Eq. A7: second order Momentum Conservation}] vanishes. To isolate the hydrodynamic fluid motion from the acoustic perturbation over time (i.e., $\partial \mathbf{v}_1 / \partial t$), we need to recall the first-order momentum conservation in Eq. \eqref{Eq. A4: first-order momentum conservation} which is rewritten as the following for convenience
\begin{equation}
\rho_{1}\frac{\partial \mathbf{v_{1}}}{\partial t} = -\frac{\rho_{1}}{\rho_{0}}\nabla p_{1} + \frac{\rho_{1}}{\rho_{0}}\left(\frac{4}{3}\mu_s+\mu_b\right)\Delta\mathbf{v_{1}}  
\label{Eq. A9: rewritten first-order momentum conservation}
\end{equation}
Substitute Eq. \eqref{Eq. A9: rewritten first-order momentum conservation} into Eq. \eqref{Eq. A7: second order Momentum Conservation}, we get
\begin{equation}
\begin{split}
-\frac{\rho_{1}}{\rho_{0}}\nabla p_{1} + \frac{\rho_{1}}{\rho_{0}}\left(\frac{4}{3}\mu_s+\mu_b\right)\Delta \mathbf{v_{1}} +  \rho_{0}\mathbf{v_{1}}\cdot\nabla\mathbf{v_{1}} \\
=-\nabla p_{2} + \mu_s \Delta \mathbf{v_{2}}
\label{Eq. A10}
\end{split}
\end{equation}

If we take the time average of the above Eq. \eqref{Eq. A10} and use the definitions of kinetic energy density $\left\langle \mathcal{K} \right\rangle = \left\langle\rho_{0}\mathbf{v_{1}}\cdot\nabla\mathbf{v_{1}}\right\rangle=\left\langle\nabla\left(1/2 \rho_{0}\mathbf{v_{1}^{2}}\right)\right\rangle$ and potential energy density $\left\langle \mathcal{U} \right\rangle=\left\langle {\rho_{1}} / {\rho_{0}}\nabla p_{1}\right\rangle=\left\langle\nabla\left({p_{1}^2} / ({2\rho_{0} c_{0}^2}) \right)\right\rangle$ (note that the first-order equation of state is applied here), the second-order momentum conservation equation can be expressed as
\begin{equation}
    \nabla\langle \mathcal{L} \rangle+\left(\frac{4 \mu_s} {3}+\mu_b \right) \left\langle \frac{\rho_1}{\rho_0} \Delta \mathbf{v}_1\right\rangle = \left\langle-\nabla p_2+\mu_s \Delta \mathbf{v}_2\right\rangle
    \label{Eq. A11: second-order momentum conservation}
\end{equation}
where the time-averaged acoustic Lagrangian is defined as $\langle \mathcal{L} \rangle = \langle \mathcal{K} - \mathcal{U} \rangle$. As noted that the acoustic Lagrangian is independent of the viscous effect and hence not affected by the wave attenuation, while the acoustic streaming is induced by the transfer of acoustic momentum to the viscous mode through wave attenuation. That is to say, the averaged acoustic Lagrangian $ \langle \mathcal{L} \rangle$ is not related to the steady acoustic streaming and can be balanced with a hydrostatic pressure gradient, having $\nabla p^*_2 =\nabla  p_2 + \langle \mathcal{L} \rangle$. \cite{lighthill1978acoustic,riaud2017influence} Hence, the second-order momentum conservation equation can be finally written as
\begin{equation}
-\nabla p^*_2+\mu_s \Delta \mathbf{v}_2 + \mathbf{F_s} = 0
    \label{Eq. A12: Stokes equation of the streaming flow}
\end{equation}
with the sole source for steady acoustic streaming is
\begin{equation}
\mathbf{F_s} = - \left(\frac{4 \mu_s} {3}+\mu_b \right) \left\langle \frac{\rho_1}{\rho_0} \Delta \mathbf{v}_1\right\rangle
    \label{Eq. A13: sole source of the streaming flow}
\end{equation}
We use the hydrodynamic Reynolds number $\mathrm{Re_{hd}}$ to characterize the relative contribution of the inertia term $\rho_{0}\left(\mathbf{v}_{2} \cdot \nabla\right) \mathbf{v}_{2}$ and the viscosity term $\mu_{s} \nabla^{2} \mathbf{v}_2$\cite{lighthill1978acoustic}:
\begin{equation}
\mathrm{Re_{hd}}=\frac{\rho_{0}{v}_2 d}{\mu_{s}}
\end{equation}
where $v_2$ is the characteristic streaming velocity of the system, $d$ is the characteristic length of the system and $d\sim 60\mu m$. The hydrodynamic Reynolds number is equal to 1 when the streaming velocity is:
\begin{equation}
     {v}_{cr}=\frac{\mu_{s}}{\rho_{0} d}=16.7 \mathrm{mm/s}
\end{equation}
When $\mathrm{Re_{hd}} \gg 1$ $\left(v_2\gg v_{cr}\right)$, the inertia term is dominant, and we use Lighthill theory to simulate the acoustic streaming, at this point, the governing equation becomes:
\begin{equation}
\rho_{0}\left({\mathbf{v_2}}\cdot\nabla\right)\mathbf{v_2}+\mu_{s}\Delta\mathbf{v_2}-\nabla p_{2}=\mathbf{F}
\end{equation}
When $\mathrm{Re_{hd}} \ll 1$ $\left({v}_2\ll{v}_{cr}\right)$, the viscosity term is dominant, and the $\rho_{0}\left(\mathbf{v_2}\cdot \nabla \right)\mathbf{v_2}$ will disappear, the governing equation will be restored to \eqref{Eq. A12: Stokes equation of the streaming flow}.
\par This makes it possible to use commercial software like COMSOL to solve the acoustic streaming problems with the source term induced by acoustic field.

\section{\label{Appendix B} Explicit expressions of source terms}
\subsection{\label{Appendix B1} Monochromatic fundamental wave}
For most applications using the acoustic streaming effect, the acoustic wave is considered as a monochromatic wave with the linear wave equation for the velocity as 
\begin{equation}
\frac{\partial^2 \mathbf{v}_1}{\partial t^2}-c_0^2 \Delta \mathbf{v}_1 = 0
    \label{Eq. B1: Wave equation}
\end{equation}
where $\partial^2 / \partial t^2 = - \omega^2$ for a single-frequency fundamental harmonic wave with $\omega = 2 \pi f$ the angular frequency. Hence, the wave equation can be simplified as $\Delta \mathbf{v}_1 = - \omega^2 \mathbf{v}_1 / c_0^2$. Insertion of the steady wave equation into the source term of Eq. \eqref{Eq. A13: sole source of the streaming flow} with the combination of the first-order equation of state, the sole source of acoustic streaming is
\begin{equation}
\begin{split}
 \mathbf{F_s}  = \left(\frac{4 \mu_s}{3}+\mu_b \right) \frac{\omega^2}{c_0^4 \rho_0}\left\langle p_1 \mathbf{v}_1\right\rangle \\
\label{Eq. B2: source for singe frequency wave}
\end{split}
\end{equation}
with the average acoustic density $\left\langle \mathbf{I} \right\rangle = \left\langle p_1 \mathbf{v}_1\right\rangle$. This formula is also given in Eq. (19) of Ref. [\onlinecite{baudoin2020acoustic}].

\subsection{\label{Appendix B2} High-order harmonic waves}
When the weak nonlinear effect induces a few high-order harmonics, e.g., up to the order of $n = 2$, the source term for the acoustic streaming given in Eq. \eqref{Eq. B2: source for singe frequency wave} will not apply which is limited to the case of single-frequency fundamental wave. Under this condition, we have to re-derive the source term based on the general expression of Eq. \eqref{Eq. A13: sole source of the streaming flow} with only two assumptions: the irrotational acoustic field ($\nabla \times \mathbf{v}_1 = 0$) and the incompressible streaming fluid field ($\nabla \cdot \mathbf{v}_2 = 0$).
By combining the first-order mass conservation equation [see Eq. \eqref{Eq. A2: first order mass conservation}] and the first-order equation of state, the relation between the acoustic pressure $p_1$ and velocity vector $\mathbf{v}_1$ can be derived as $\nabla p_1 /(\rho_0 c_0^2) = \Delta \mathbf{v}_1$ with vector identity $\nabla\nabla\cdot\mathbf{v_{1}}=\nabla^{2}\mathbf{v_{1}}+\nabla\times\nabla\times\mathbf{v_{1}}$. By insertion of this relation into the general expression of body force $\mathbf{F_s}$ in Eq. \eqref{Eq. A13: sole source of the streaming flow}, one can derive the following source form including only the acoustic pressure: 
\begin{equation}
\mathbf{F_s} = \left(\frac{4}{3} \mu_s+\mu_b\right) \left\langle\frac{{p}_{1}}{c_{0}^{4} \rho_{0}^{2}} \nabla \frac{\partial {p}_{1}}{\partial t}\right\rangle 
\label{Eq. B3: source in terms of only pressure}
\end{equation}
This form is helpful to deal with acoustic streaming problems if the acoustic field with high-order harmonics is solved with either analytical or numerical methods. 

Consider the acoustic pressure form including low-order harmonics\cite{Du1986Harmonic}:
\begin{equation}
\begin{split}
p_1 & =P_{1} \sin \left[\omega\left(t- \frac{z}{c_{0}}\right)\right] + P_{2} \sin\left[2\omega\left(t-\frac{z}{c_{0}}\right)\right] \\ 
& = \sum_{n=1}^{2} P_{n} \sin \left[n \omega\left(t-\frac{z}{c_{0}}\right)\right]
\end{split}
\end{equation}
\par Substituting the above equation into Equation \eqref{Eq. B3: source in terms of only pressure}, the following form of body force can be obtained:
\begin{equation}
F_{z}^{h}=\frac{\alpha}{\rho_{0} c_{0}^{2}} \sum_{n=1}^{2}\left\langle n^{2} P_{n}^{2}\right\rangle
\end{equation}
The orthogonality of trigonometric functions is used in the derivation of the above formula, that is $1/T \int_0^T  [sin\left(n \omega t\right)sin\left(m \omega t\right)] dt$ = 0 when $n\neq m$ and $T = 2\pi / \omega$ is the period for the fundamental harmonics.

\subsection{\label{Appendix B3} Pressure fields of high-order harmonic waves by Du \& Breazeale}
By using the perturbation method to solve the nonlinear wave equation (first proposed by Kuznetsov \cite{1971Equation}), we can obtain the following fundamental quasilinear solution (see Eq.(4) in see Ref. [\onlinecite{Du1986Harmonic}] with $z/z_{0}$$\ll$1 and $\gamma$$\approx$$\pi$/2):
\begin{equation}
P_{1}(f \mid z)=p_{am}\frac{e^{-\alpha z}}{\sqrt{1+\left(z/z_{0}\right)^2}}\exp\left({-\frac{R^{2}/a^{2}}{1+\left(z/z_{0}\right)^{2}}}\right)
\label{Eq.B6}
\end{equation}
where $z_{0}=k a^{2} / 2$. It is noted that the near-field condition is applicable 
($z \ll z_{0}$) when the channel height is small. In this case, the diffraction effect is very weak, and therefore the acoustic pressure in Eq. \eqref{Eq.B6} can be simplified as
\begin{equation}
P_{1}(f \mid z)=p_{am} e^{-\alpha z} e^{\left(-R^{2} / a^{2}\right)} \label{simplify}
\end{equation}

Noted that in Blackstock's study \cite{Blackstock1964Thermoviscous}, the extra attenuation (EXDB) is very small (The EXDB refers to the loss beyond the normal small signal attenuation $e^{-\alpha z}$) when the Goldberg number $\Gamma $ $\sim$ 1. 
Under this condition, low-order harmonics will be generated, while no shock waves will be formed. 
With the help of the Hankel transform, $P_{2}$ can be calculated by following the work of Du \& Breazeale [see Eqs. (11) and (12) in Ref. [\onlinecite{Du1986Harmonic}]):
\begin{equation}
\begin{split}
P_{2}(f \mid z)=\frac{p_{am} D_0 e^{-4 \alpha z}}{4\sqrt{1+(z/z_0)^2}}\exp{\left(-\frac{2R^2/a^2}{1+(z/z_0)^2}\right)} \\ \times \sqrt{H_{2}^2+F_{2}^2}
\end{split}
\label{Eq.20}
\end{equation}
with
\begin{equation}
    H_{2}=\int_{0}^{\sigma} \frac{e^{2\alpha z_{0}\sigma'}}{\sqrt{1+\left(z/z_0\right)^2}}\cos\left[\tan^{-1}\left(\sigma'\right)\right]d\sigma'
\end{equation}
and 
\begin{equation}
     F_{2}=\int_{0}^{\sigma} \frac{e^{2\alpha z_{0}\sigma'}}{\sqrt{1+\left(z/z_0\right)^2}}\sin\left[\tan^{-1}\left(\sigma'\right)\right]d\sigma'
\end{equation}
where $\sigma=z/z_0$, and $D_0= {2z_0}/{L_s}$. Considering the near-field condition $z \ll z_{0}$, we can further simplify the second harmonic acoustic pressure Eq. \eqref{Eq.20} as: 
\begin{equation}
P_{2}(f \mid z)=\frac{p_{am}^{2} k \beta e^{-4 \alpha z}}{4 \alpha \rho_{0} c_{0}^{2}} \exp \left(-2 R^{2} / a^{2}\right)\left(e^{2 \alpha z}-1\right) \label{secondharmonic}
\end{equation}
Note that the solutions for higher harmonics up to the fourth order are also listed in the Appendix of Ref. [\onlinecite{Du1986Harmonic}]. 


\renewcommand\refname{Reference}
\bibliography{main}        

\end{document}